\documentclass[structabstract]{aa}  
\usepackage{graphicx}
\usepackage{txfonts}
\begin{document}
\title{Identification of the TW Hya Association member 2M1235$-$39:
  a tertiary component of the HR 4796 system \\ 
  {\it Research Note}}


   \author{Joel H. Kastner
          \inst{1,2}
          \and
          B. Zuckerman\inst{3,4}
          \and
          M. Bessell\inst{5}
          }

   \institute{Laboratoire
  d'Astrophysique de Grenoble, Universit\'e Joseph Fourier --- CNRS,
  BP 53, 38041 Grenoble Cedex, France\\
              \email{joel.kastner@obs.ujf-grenoble.fr}
         \and
         Chester F. Carlson Center for Imaging
  Science, Rochester Institute of Technology, 54 Lomb Memorial Dr.,
  Rochester, NY 14623 USA
         \and
             Dept.\ of Physics \& Astronomy, University of California, Los
  Angeles 90095 USA 
         \and 
             UCLA Center for Astrobiology, University of California, Los
  Angeles 90095 USA
         \and
             Research School of Astronomy and Astrophysics, The
             Australian National University, Weston, ACT 2611, Australia
             }

\titlerunning{Identification of TWA Member 2M1235$-$39}
\authorrunning{Kastner, Zuckerman, \& Bessell}
   \date{Received ...; accepted ...}

 
  \abstract
{}
{We seek to determine whether the late-type star 2MASS
  J12354893$-$3950245 (2M1235$-$39) is a
  member of the TW Hya Association (TWA), a hypothesis suggested by
  its association with a bright X-ray 
  source detected serendipitously by ROSAT and XMM-Newton and its
  ($\sim3'$) proximity to the well-studied (A+M binary) system HR 4796.}
  {We used optical spectroscopy to establish the Li and
    H$\alpha$ line strengths of 2M1235$-$39, and determined its proper
    motion via optical imaging. We also considered its X-ray and near-IR
  fluxes relative to the M star HR 4796B.}
{The optical spectrum of 2M1235$-$39 displays strong Li absorption and
  H$\alpha$ emission (equivalent widths of 630 m\AA\ and $-6.7$ \AA,
  respectively). Comparison of the spectrum with that of a nearby
  field star, along with the DENIS catalog $IJK$ magnitudes, indicates
  the spectral type of 2M1235$-$39 is M4.5. We measure a proper motion
  for 2M1235$-$39 that agrees, within the errors, with that of HR
  4796.}
  {The Li absorption and H$\alpha$ emission line strengths of
    2M1235$-$39, its near-IR and X-ray fluxes, and its proper motion
    all indicate that 2M1235$-$39 is a TWA member. Most likely this
    star is a wide (13,500 AU) separation, low-mass, tertiary
    component of the HR 4796 system.}


   \maketitle
%

\section{Introduction}

As of little more than a decade ago, astronomers were almost
oblivious to the presence of low-mass, pre-main sequence 
stars within $\sim100$ pc of Earth. The intervening years have
seen the identification of a few hundred such stars, with ages ranging
from 8 to 100 Myr, as part of numerous post-T Tauri
associations (Zuckerman \& Song 2004, hereafter ZS04, and references
therein; Torres et al.\ 2006, 2008). Perhaps the greatest
excitement associated with the recognition of the existence of nearby
young stars has been the opportunity to study, at
close range, the evolution of youthful planetary systems, via direct
thermal imaging of warm massive planets (e.g., Chauvin et al.\ 2004; Song et
al.\ 2006) and via imaging and spectroscopy of debris disks (e.g.,
Rebull et al.\ 2008 and references therein). Young, local stellar
groups also afford unique insight into the early evolution
of low-mass stars and ultracool dwarfs (e.g., Looper et
al.\ 2007; Cruz et al.\ 2008; and references therein).

The difficulty inherent in identifying young stars and young star
groups near Earth reflects the fact that such groups are spread over
large areas of the sky (ZS04). Furthermore, while the local young
groups are usually ``spearheaded'' by a handful of well-studied,
individual systems that feature, e.g., strong H$\alpha$ emission,
enormous IR excesses, and/or easily imaged debris disks (TW Hya and
$\beta$ Pic being cases in point), the vast majority of nearby young
stars are otherwise unremarkable late-type (K through M) dwarfs that
do not stand out or even turn up in optical emission-line or
far-infrared (e.g., IRAS) surveys.

However, all $\sim10$--100 Myr-old stars of types F through M are at
or near the peaks of their lives in terms of their X-ray luminosities
relative to bolometric (with ``saturated'' values of $L_X/L_{\rm bol}
\sim 10^{-3}$, ZS04 [their Fig.~4]; see also Kastner et al.\ 1997 and
Preibisch \& Feigelson 2005). Hence, X-ray point source catalogs, in
tandem with recently released, comprehensive catalogs of distances and
proper motions of stars in the solar neighborhood, have served as the
main resources with which to isolate stars that are likely nearby and
young. Followup optical spectroscopy and/or imaging then readily
confirms (or refutes) membership in the ``nearby young star club,''
via determination of surface Li abundances and relative ($UVW$)
Galactic space motions.

Here, we demonstrate that serendipitous XMM-Newton and ROSAT X-ray
detections of 2MASS J12354893$-$3950245 (hereafter 2M1235$-$39),
combined with its optical spectrum and proper motion, establishes 
this star as a member of the quintessential local young star group,
the TW Hya Association (TWA; Kastner et al.\ 1997; Webb et al.\ 1999;
Zuckerman et al.\ 2001). Indeed, 2M1235$-$39
is, likely, the tertiary component of the well-studied HR 4796 (A+M
star) binary system (Jura et al.\ 1993; Stauffer et al.\ 1995), which
is designated TWA 11.

\section{Observations and Results}

\subsection{Serendipitous X-ray detections of 2M1235$-$39}

\subsubsection{ROSAT}

No X-ray sources are associated with 2M1235$-$39 in the ROSAT
All-Sky Survey Bright or Faint Source Catalogs. However, the HR 4796
system was the subject of a pointed 41 ks ROSAT HRI observation (Jura
et al.\ 1998). That observation resulted in detections of HR 4796B and
a source within $5''$ of the position of 2M1235$-$39 (ROSAT HRI count
rates\footnote{The HRI X-ray count rates listed here were obtained
  from the HEASARC ROSAT archive; the count rate for HR 4796B is
  consistent with that determined by Jura et al.\ (1998).}
23.7$\pm$0.8 ks$^{-1}$ and 8.4$\pm$0.5 ks$^{-1}$, respectively). The
ROSAT/HRI detection of HR 4796B was used by Jura et al.\ to establish
that the X-ray source was centered on this M star, rather than on the
primary (A-type) star. The detection of 2M1235$-$39 was not noted by
these authors, however\footnote{Indeed, the recognition that HR 4796
  is a member of the TWA (Webb et al.\ 1999) came subsequent to
  publication of Jura et al.\ (1998).}.

\subsubsection{XMM-Newton}

The serendipitous XMM-Newton detection of 2M1235$-$39 is summarized in
Lopez-Santiago et al.\ (2007), who analyzed 58 bright (EPIC-MOS2
0.5--4.5 keV count rate $>10$ ks$^{-1}$) X-ray sources with stellar
counterparts that were detected in the XMM Bright Serendiptious Survey
(XBSS). In the Lopez-Santiago et al.\ study, 2M1235$-$39 was
identified as a star of M4 spectral type, where this spectral type is
an estimate based its 2MASS colors.  A better defined color-index
for spectral typing of mid-M-type stars is I$-$K which, from DENIS, 
indicates a spectral type of M4.5.

Given the relative paucity of serendipitous XMM sources as bright as
that associated with 2M1235$-$39 (Lopez-Santiago et al.\ 2007), the
proximity of the 2M1235$-$39 X-ray source to HR 4796AB (angular
displacement of $\sim3'$) as well as the similarity of the X-ray
fluxes and temperatures determined for the 2M1235$-$39 and HR 4796B
sources by Lopez-Santiago et al.\ (2007) suggested to us that 2M1235$-$39
might be an as-yet unrecognized member of the TWA and, perhaps, a
widely separated companion to HR 4796AB.

\subsection{Optical spectroscopy of 2M1235$-$39 and reference star}

\begin{figure}[htb]
\begin{center}
\includegraphics[scale=0.4,angle=0]{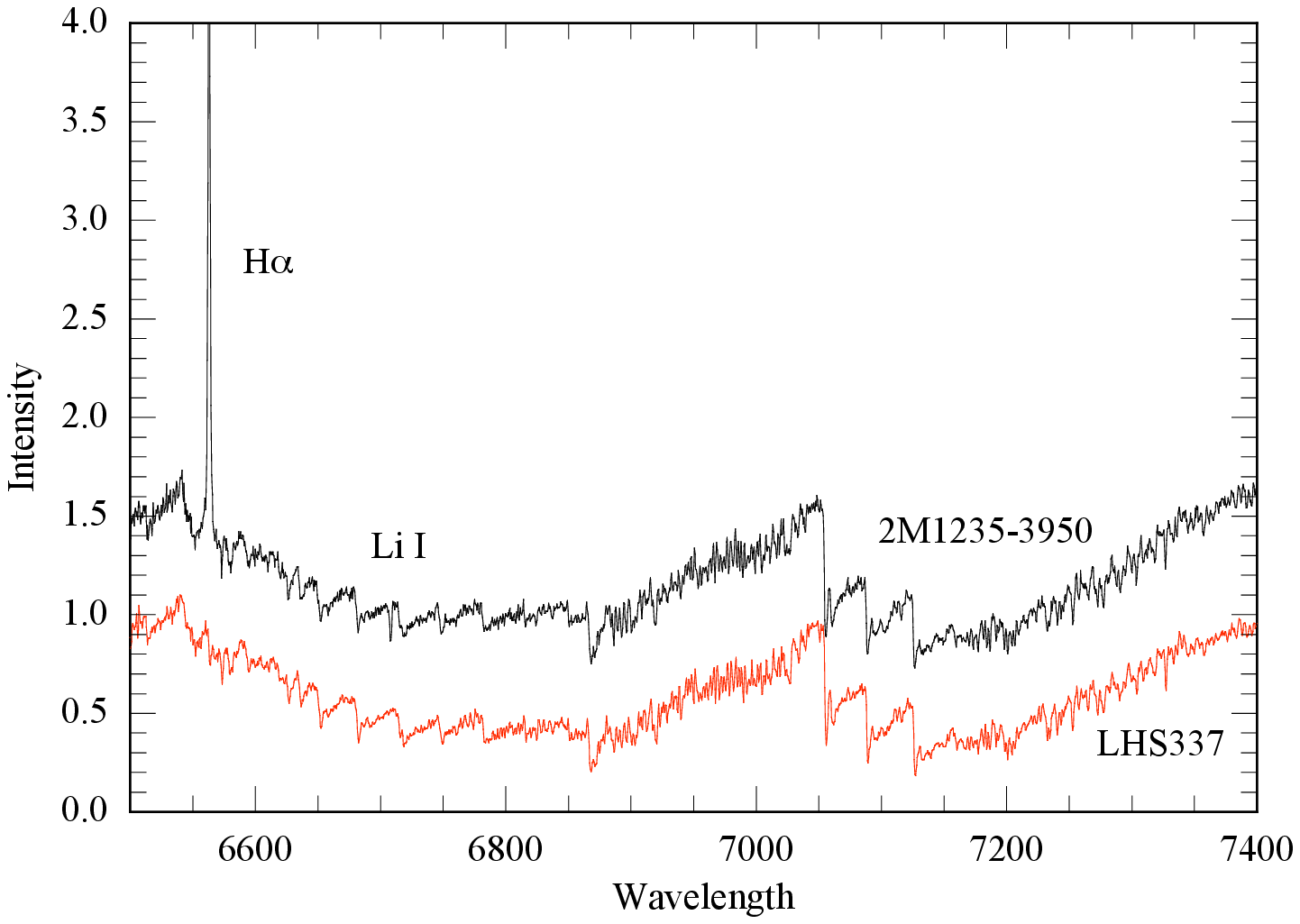}
\includegraphics[scale=0.4,angle=0]{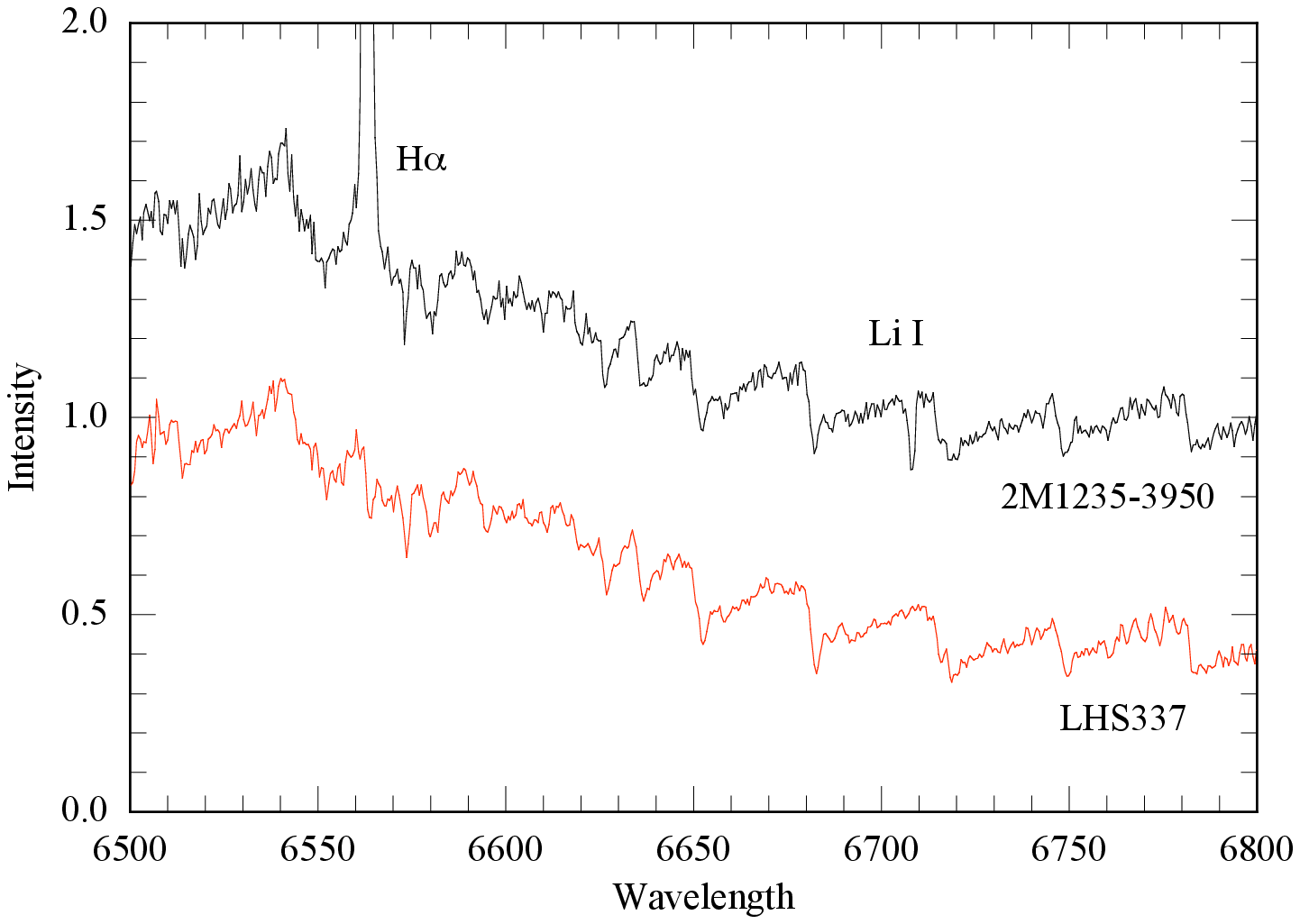}
\end{center}
\caption{Spectra of 2M1235$-$39 and a field star
  (LHS 337) of M4 spectral type, obtained with the Double
  Beam Spectrograph on the 2.3 m Siding Spring Observatory telescope
  (top: entire spectral range, bottom: closeup of the H$\alpha$ and Li
  spectral region). Note the deep Li absorption and strong
  H$\alpha$ emission in the 2M1235$-$39 spectrum. These spectral
  features are typical of the 8 Myr-old stars of the TWA.}
\label{fig:TWA11Cspec}
\end{figure}

To test the foregoing hypothesis, we obtained moderate-resolution
($R\sim 1.16$ \AA) red (6500--7500\AA\ region) spectra of 2M1235$-$39 and
the nearby, M4 spectral type field star LHS 337 with the Double Beam
Spectrograph (DBS) grating spectrometer on the 2.3 m Siding Spring
Observatory (SSO) telescope. Results for the spectral range covering
the $\lambda$6563 H$\alpha$ and $\lambda$6708 Li {\sc i} lines are
displayed in Fig.\ 1. As expected for a member of the TWA, 2M1235$-$39
displays strong Li absorption and H$\alpha$ emission, with respective
equivalent widths 630 m\AA\ and $-6.66$ \AA. The lithium measurement confirms 
the youth of this star, as M
dwarfs do not retain measureable abundances of Li beyond $\sim30$ Myr
(see ZS04 and references therein). Furthermore, the spectral
similarity of 2M1235$-$39 to LHS 337 (apart from the Li and H$\alpha$
lines) confirms that its spectral subtype is consistent with M4.5.

\subsection{Proper motion of 2M1235$-$39}

To ascertain the proper motion of 2M1235$-$39, we obtained a set of
$I$ band images of the field with a 512$\times$512 CCD camera on
the 1 m telescope of the University of Tasmania Mt.\ Canopus
Observatory on 10 June 2008 (epoch 2008.44). The 
camera pixel scale is 0.434$\pm$0.003 arcsec pix$^{-1}$.  With these
images and the ESO 1982.37 epoch archival digitized sky image of the
2M1235$-$39 field, the proper motion of 2M1235$-$39 was measured
using the IRAF task \verb+geomap+ (D. Rodriguez 2008, personal comm.).
The resulting proper motion is $-49.6\pm3$ and $-25.1\pm3$ mas
yr$^{-1}$ in RA and dec, respectively.  Within the respective errors,
these proper motions agree with an average of those measured for HR
4796A in the Hipparchos, TYCHO, and PPM catalogs, i.e., 
($-55.9$, $-24$), ($-53.3, -21.2$), and ($-46$, $-18$) mas yr$^{-1}$,
respectively.

\section{Discussion}

The strong Li absorption measured for 2M1235$-$39 is compatible
with TWA membership (e.g., Kastner et al.\ 1997; Webb et al.\ 1999;
Zuckerman et al.\ 2001; Song et al.\ 2003).  Specifically, the
previously established member stars of the TWA that are of M spectral
type have 6708 \AA\ Li absorption line EWs in the range 360--650
m\AA. In addition, its H$\alpha$ emission-line strength is
near the median for M stars in the TWA. 

Adopting a conversion factor of $1.1\times10^{-11}$ erg cm$^{-2}$
count$^{-1}$ (Jura et al.\ 1998), the respective ROSAT/HRI (0.1--2.4
keV) X-ray fluxes are $9.2\times10^{-14}$ erg cm$^{-2}$ s$^{-1}$ and
$2.8\times10^{-13}$ erg cm$^{-2}$ s$^{-1}$ for the 2M1235$-$39 and HR
4796B X-ray sources, respectively, whereas Lopez-Santiago et al.\
(2007) derive respective intrinsic (0.5--10 keV) X-ray fluxes of $F_X
= 2.2\times10^{-13}$ erg cm$^{-2}$ s$^{-1}$ and $F_X =
1.4\times10^{-12}$ erg cm$^{-2}$ s$^{-1}$ from spectral modeling of
the XMM-Newton EPIC CCD data. As the X-ray spectra of both sources
display essentially no absorption (Lopez-Santiago et al.\ 2007), the latter
intrinsic fluxes also well represent their observed 0.5--10 keV fluxes
as measured by XMM. The discrepancies between the fluxes of the
2M1235$-$39 and HR 4796B X-ray sources as measured by XMM/EPIC vs.\
ROSAT/HRI --- i.e., a factor $\sim2.5$ for 2M1235$-$39 and $\sim5$ for
HR 4796B --- are likely due in large part to the superior hard X-ray
sensitivity of XMM/EPIC, since each source displays not only a soft
($kT \sim 0.3$ keV) but a hard ($kT \sim 1.0$ keV) X-ray component in
XMM/EPIC spectra (Lopez-Santiago et al.).  It is also likely that
these sources display X-ray source variability typical of pre-MS
stars. In any event, the flux of the X-ray source associated with
2M1235$-$39 is compatible with TWA membership, given its mid-M
spectral type (e.g., Kastner et al.\ 1997).

The proper motion of 2M1235$-$39 is also consistent with TWA
membership and, as noted in Section 2.3, agrees within the errors with
that of the HR 4796AB (= TWA 11AB = CD$-39$ 7717AB\footnote{The
  identification of an ASCA X-ray field source as ``CD$-$39 7717C'' in
  Turner et al (1997) appears to be spurious. This source is, most
  likely, X-ray emission from HR 4796B or (perhaps) a combination of
  flux from HR 4796B and 2M1235$-$39, since the (poorly-determined)
  ASCA position for this source --- while closer to HR 4796B --- does
  not correspond well to the position of either star. }) binary
system.  If 2M1235$-$39 is at the revised Hipparcos-measured distance
of HR 4796A, 78.5 pc (van Leeuwen 2007), then the absolute K magnitude
of 2M1235$-$39 is 4.5, based on the 2MASS and DENIS catalogs.  This
M$_K$ places 2M1235$-$39 within or perhaps slightly above the locus of
5--8 Myr-old stars in the TWA and the $\eta$ Cha cluster, very near
the position of a binary M4.5 system in $\eta$ Cha (e.g., Fig.\ 2 in
Song et al.\ 2003 and Fig.\ 2 in ZS04). Its similarity to the latter
system suggests that 2M1235$-$39 may be a close binary.  Given these
considerations and the very low areal density of known TWA members,
most likely 2M1235$-$39 (``TWA 11C'') and TWA 11AB are physically
bound, with a projected separation of 13,500 AU. This would make TWA
11AB and TWA 11C the widest known binary in the TWA by a large margin
(the previously identified TWA binary systems all have separations
less than $\sim15''$, or $\sim750$ AU).

\section{Conclusions}

All measured properties of 2M1235$-$39 presented here --- its Li
absorption and H$\alpha$ emission line strengths, its near-IR and
X-ray fluxes, and its proper motion --- are compatible with TWA
membership. Based on these results and on the similarity of its
common proper motion to that of HR 4796A, we conclude that 2M1235$-$39
is most likely a wide (13,500 AU) separation, low-mass, tertiary
component of the HR 4796 system.

\acknowledgements{The authors are grateful to John Greenhill for
  obtaining the set of $I$ band images and David Rodriguez and Jay
  Farihi for analyzing these current epoch images of 2M1235$-$39. We
  thank Thierry Forveille for alerting us to the revised Hipparcos
  distance to HR 4796. J.H.K. thanks the staff of the Laboratoire
  d'Astrophysique de Grenoble for their support and hospitality during
  his yearlong sabbatical visit to that institution.  This research
  was partially supported by a NASA grant to UCLA.}

\end{document}